# Integrated photonic computing: towards high-dimensional information processing

Ji Qin[1,†], Zhi-Kai Pong[1,†], Xuke Qiu[1,†], Liangyu Deng[1,†], Runchen Zhang[1], Yunqi Zhang[1], Jinge Guo[1], Yifei Ma[1], Zimo Zhao[1], Yuanxing Shen[2], Patrick Salter[1], Martin Booth[1], Stephen Morris[1], Honghui He[2], Min Gu[3,4], Bowei Dong[5,*] and Chao He[1,*]

[1]*Department of Engineering Science, University of Oxford, Parks Road, Oxford, OX1 3PJ, UK*
[2]*Guangdong Research Center of Polarization Imaging and Measurement Engineering Technology, Institute of Biopharmaceutical and Health Engineering, Tsinghua Shenzhen International Graduate School, Tsinghua University, Shenzhen 518055, China*
[3]*School of Artificial Intelligence Science and Technology, University of Shanghai for Science and Technology, Shanghai 200093, China*
[4]*Institute of Photonic Chips, University of Shanghai for Science and Technology, Shanghai 200093, China*
[5]*Institute of Microelectronics (IME), Agency for Science, Technology and Research (A*STAR), Singapore, Republic of Singapore*

[†]*These authors contributed equally to this work*
[*]*Corresponding authors: chao.he@eng.ox.ac.uk; dong_bowei@a-star.edu.sg*

## Abstract

**The rapid growth of artificial intelligence, coupled with the slowing of Moore's law, is straining computing infrastructure, as CMOS electronics face inherent limits in bandwidth, energy efficiency, and parallelism. Integrated photonic computing encodes and processes information using the phase, amplitude, spatial modes, wavelength channels, and polarisation of guided optical fields, offering a scalable and energy-efficient route beyond charge-based signalling. Here, we review on-chip photonic computing, emphasising the progression from low-dimensional to high-dimensional architectures. At the foundational level, low-dimensional approaches manipulate the phase and amplitude of guided light through Mach-Zehnder interferometers, diffractive structures, microring resonators, and absorptive elements, forming a programmable basis for optical matrix-vector multiplication. Crucially, high-dimensional architectures exploit spatial modes and wavelength channels to carry multiple independent data streams through a single waveguide, achieving higher throughput with moderate hardware overhead. Practical deployment, however, demands more than device innovation. We examine how system-level techniques, from time-wavelength interleaving to hardware-aware training, address energy efficiency, precision, and algorithm-hardware co-design. Five challenges nevertheless remain: electro-optic conversion efficiency, computing parallelism, spatial integration, reconfigurability, and robustness. We highlight emerging topological structures, such as optical skyrmions, as a promising route to fault-tolerant, topologically protected encoding that exploits the largely untapped polarisation degree of freedom. We argue that, by embracing the higher dimensionality of light, photonic computing can offer not merely an incremental improvement but a new paradigm for high-performance, energy-efficient information processing.**

## 1. Introduction

The rapid advancement of artificial intelligence (AI), driven by ever-larger models and data-intensive workloads, is placing unprecedented demands on computing infrastructure. Graphics processing unit (GPU)-based systems, still fundamentally reliant on complementary metal-oxide-semiconductor (CMOS) electronics, are struggling to deliver sufficient performance within acceptable power and thermal limits [1–10]. Moreover, with Moore's Law slowing and Dennard scaling breaking down [11–13], the gains from continued transistor miniaturisation have become marginal [11,14]. Most critically, the charge-based and largely sequential nature of electronic circuits imposes inherent limits on both parallelism and energy efficiency. These physical and architectural constraints indicate that merely scaling existing hardware is no longer a sufficient path forward [15–19]. Meeting the demands of next-generation AI will instead require fundamentally new computing paradigms that extend beyond the boundaries of conventional electronics.

Photonic computing, which performs computation with photons rather than electrons, offers a promising route to overcome these limitations [20–22]. By exploiting the physical properties of light, photonic systems can reach superior performance in capacity, density, efficiency, and latency compared with their electronic counterparts [21,23–28]. Recent progress has demonstrated key photonic computing modules, underscoring the potential of integrated optics to reshape AI acceleration [29–33]. Among these, a particularly compelling direction is the development of all-optical neural networks (ONNs), which aim to reduce reliance on electronic processing by executing key computations optically [34–37]. ONNs exploit three key advantages of light: (1) ultrawide bandwidth—from tens of gigahertz to hundreds of terahertz—and intrinsic parallelism, which together enable high-throughput processing critical for accelerating large-scale inference and training [38,39]; (2) minimal propagation loss (thin-film lithium-niobate waveguides, for instance, achieve losses ~0.06 dB per centimetre at visible wavelengths [40]), which reduces thermal overhead [41]; and (3) multidimensional encoding through polarisation and spatial modes [42,43], enabling scalable, non-interfering, parallel computations that can significantly increase computational density.

To implement these capabilities, photonic computing platforms can be broadly categorised into free-space systems and integrated on-chip implementations [44]. This review focuses on the integrated approach, which leverages mature CMOS-compatible fabrication processes to achieve large-scale integration of thousands of optical components on a single chip. Our discussion centres on matrix-vector multiplication (MVM), a core operation in AI workloads and a natural fit for optical acceleration. Readers interested in free-space systems are referred to the relevant literature [22,45]. A central theme of this review is the emerging role of high-dimensional computing, which exploits additional degrees of freedom of light to enhance throughput and processing density.

Within this integrated photonic computing paradigm, approaches can be viewed along a progression of exploited optical degrees of freedom. At the foundational level, low-dimensional computing directly modulates phase and amplitude—the most commonly exploited degrees of freedom. Phase-based schemes shape the optical wavefront through devices such as Mach-Zehnder interferometers (MZIs) and diffractive structures, while amplitude-based designs control light intensity using microring resonators (MRRs) or absorptive elements. Fig. 1a summarises these representative building blocks. Extending this foundation, high-dimensional computing additionally leverages spatial modes via mode-division multiplexing (MDM) and spectral channels via wavelength-division multiplexing (WDM), further parallelising and scaling computation while retaining the same underlying modulation principles. Beyond device-level implementations, system-level strategies—including time-wavelength interleaving, amortisation, coherent detection, efficient learning methods, and hardware-aware training—address practical challenges in input encoding, output detection, and algorithmic co-design. This progression—from low-dimensional modulation, through high-dimensional parallelisation, to system-level co-design—serves as the organisational basis for the remainder of this review.

As illustrated in Fig. 1b, we organise our discussion as follows. Having established the case for photonic computing in this Introduction, Section 2 introduces the optical properties of light that underpin optical computation. Section 3 examines implementation strategies across three levels: Section 3.1 covers low-dimensional computing; Section 3.2 addresses high-dimensional computing; and Section 3.3 discusses system-level optimisations. Section 4 identifies five major challenges limiting scalability and practicality—electro-optic efficiency, computing parallelism, spatial integration, reconfigurability, and robustness—and outlines future directions. Throughout, we place particular emphasis on high-dimensional computing architectures, which leverage the intrinsic multidimensionality of light to enhance both throughput and processing density.

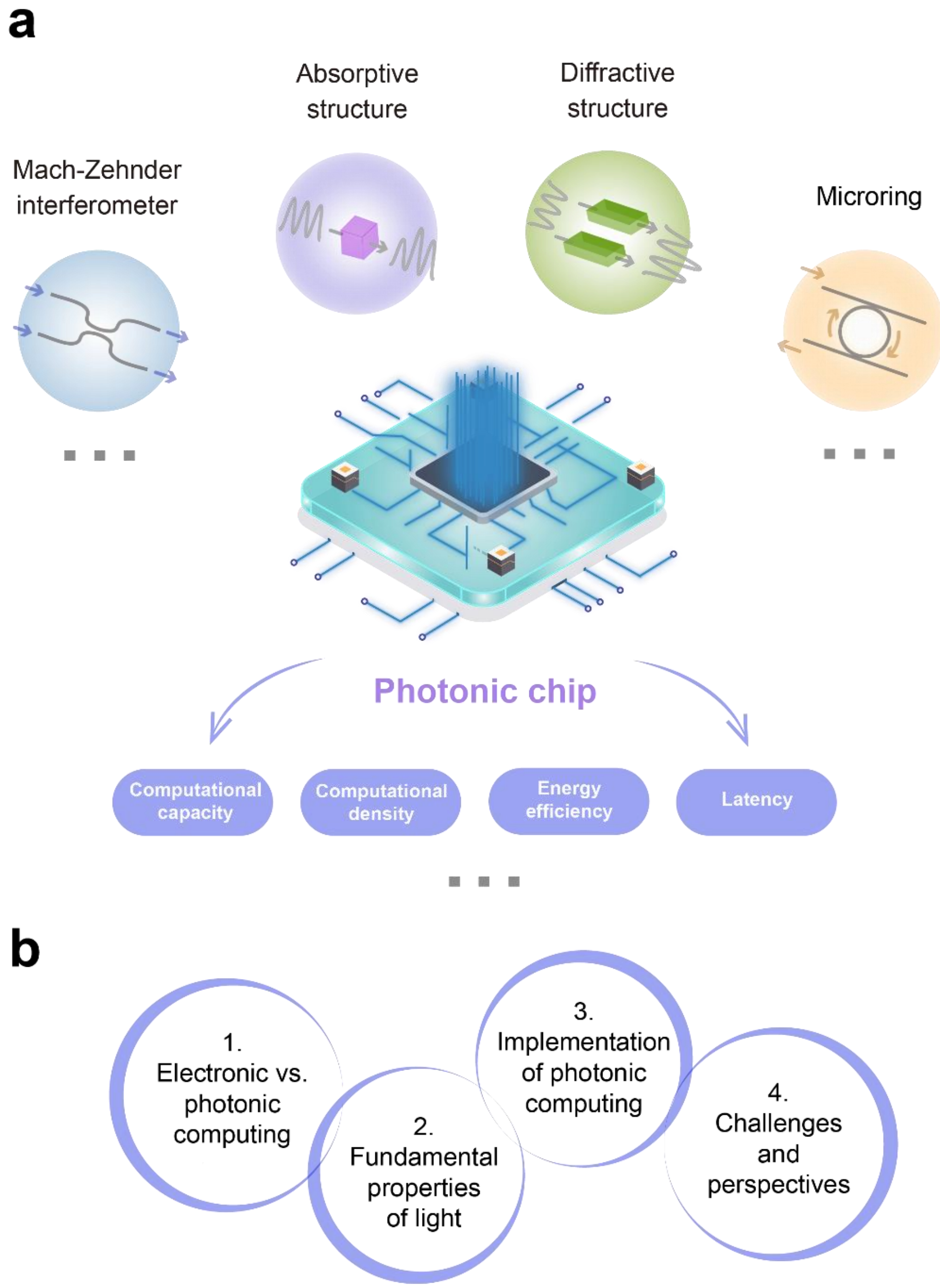

**Fig. 1 Overview of the advantages of optical chips and the structure of this review.** (a) Optical computing chips highlight superior computational capacity, computational density, energy efficiency, and latency through building blocks such as the MZI, absorptive structures, diffractive structures, and MRRs. (b) Roadmap of this review, illustrating the logical progression from the comparison between electronic and photonic computing, to the fundamental properties of light, the implementation of photonic computing, and finally the challenges and perspectives for future development.

## 2. Fundamental properties of light

The optical properties of light most directly exploited in photonic computing include phase, amplitude, polarisation, spatial mode structure, and spectral content, which together form the basis for optical information encoding and computation. Depending on whether optical phase relationships are preserved and accessible, these properties can be exploited through coherent strategies, which manipulate the complex optical field, or incoherent (intensity-based) strategies that operate primarily on optical power; both paradigms feature prominently in current photonic computing architectures (see Sections 3.1-3.3). Beyond the choice of coherent or incoherent operation, temporal structuring, while not unique to light, can further complement these optical properties as a system-level multiplexing axis, particularly when combined with spectral channels in photonic hardware. These properties provide distinct channels through which information can be represented, controlled, and processed. Together, they underpin both low-dimensional operations and high-dimensional computing schemes. To illustrate how these properties enter the mathematical description, we adopt a simplified notation for one guided optical-field component propagating along $z$ : $E = A(S)\vec{\mathbf{e}}\cos(kz + \omega t + \phi)$ , where $S = (x, y)$ denotes the transverse spatial coordinates in the plane perpendicular to the propagation direction $z$, and $A(S)$ denotes the transverse mode-profile amplitude as a function of $S$ ; $\vec{\mathbf{e}}$ is the unit polarisation vector; $k$ the wave vector; $\omega$ the angular frequency; $t$ time; and $\phi$ the phase offset. A general optical field may be represented as a superposition of such terms over spatial modes, spectral components, and polarisation states. We note that this notation is simplified: when phase relationships are relevant, a rigorous treatment would use complex-exponential (phasor) form for a coherent, fully polarised field component and, for free-space beams such as Laguerre-Gaussian modes, would include the z-dependent evolution of the transverse profile. Incoherent or intensity-based implementations are more naturally described in terms of optical power, intensity, or correlation functions, while partially polarised or depolarised fields require statistical descriptions, such as coherence-matrix or Stokes-Mueller formalisms. For guided modes in uniform waveguides, the primary setting of this review, the transverse profile is $z$ -independent and the expression above captures the essential controllable optical quantities. Fig. 2 illustrates how these optical properties map onto the low-dimensional and high-dimensional computing paradigms. Understanding how these optical properties contribute to computation is therefore key to designing effective photonic computing architectures.

As introduced in Section 1, photonic computing can be viewed as a progression from low-dimensional computing (operating on phase and amplitude) to high-dimensional computing, which additionally exploits spatial modes and spectral channels. This review places particular emphasis on the latter: by exploiting MDM and WDM, high-dimensional approaches can substantially increase throughput and computational density, although additional multiplexing and demultiplexing components may be required. Notably, polarisation $\vec{\mathbf{e}}$ remains relatively underexplored in this context. Although spatial-mode techniques can implicitly involve polarisation degrees of freedom, dedicated polarisation-based encoding offers further opportunities to expand the dimensionality of photonic computing (see Section 4).

In the sections that follow, we examine representative photonic computing architectures, analyse their operating principles, and discuss how multidimensional encoding strategies build upon low-dimensional mechanisms to address the growing computational demands of modern AI systems.

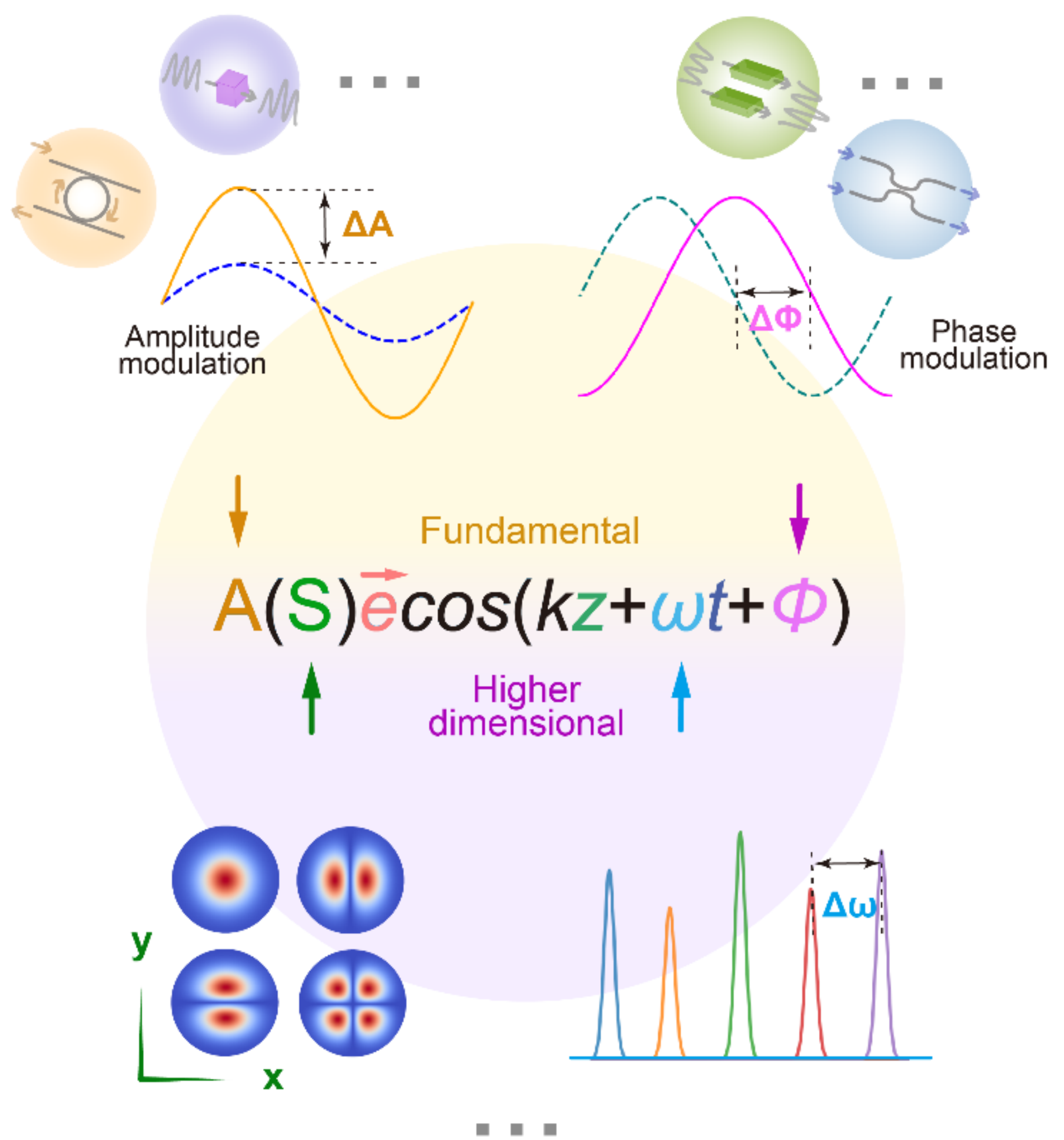

**Fig. 2 Categorisation and implementation approaches in photonic computing.** For compactness, the illustration uses a simplified notation for one optical-field component, $A(S)\vec{\mathbf{e}}\cos(kz+\omega t+\phi)$, to label controllable optical quantities without restricting the methods to coherent or single-mode operation. Here, $S=(x,y)$ denotes the transverse spatial coordinates in the plane perpendicular to the propagation direction $z$; $A(S)$ represents the transverse mode-profile amplitude; $\vec{\mathbf{e}}$ is a unit polarisation vector; and $(kz+\omega t+\phi)$ denotes the phase factor. Specific implementations may operate coherently or incoherently, depending on whether phase relationships are preserved and accessed. Approaches range from low-dimensional computing, which manipulates phase and amplitude, to high-dimensional computing, which additionally exploits spatial modes and spectral channels.

## 3. Implementation of photonic computing

This section outlines how the optical degrees of freedom introduced in Section 2 are harnessed for practical computing, bridging physical principles with functional architectures at both the device and system levels. Following the framework established in Section 1, we examine low-dimensional computing approaches in Section 3.1, high-dimensional computing in Section 3.2, and system-level optimisations in Section 3.3.

### 3.1 Low-dimensional photonic computing: phase and amplitude modulation

Low-dimensional photonic computing directly modulates the phase or amplitude of light—two commonly exploited degrees of freedom that underpin most integrated optical designs. The following sections detail how each of the four device classes introduced in Section 1—MZIs, diffractive structures, MRRs, and absorptive elements—is implemented in hardware.

#### Phase modulation: Mach-Zehnder interferometers

The MZI is one of the most widely used phase-control devices in integrated optical systems [46, 47]. A standard MZI consists of two directional couplers connected by parallel waveguides, with phase shifters integrated into one or both arms to tune the phase delay (Fig. 3a(i)). By locally modifying the refractive index, MZIs can be dynamically reconfigured in response to electrical or thermal input, making them well-suited for programmable photonic computing. When deployed in mesh networks, MZIs can implement a broad range of analog linear transformations, including MVM [48, 49].

Building on this foundation, self-configuring MZI meshes have been demonstrated that perform arbitrary linear optical transformations using only local feedback (Fig. 3a(ii)) [50, 51], and subsequent work has realised on-chip neural networks for energy-efficient AI inference (Fig. 3a(iii)) [48, 52, 53]. More recently, large-scale MZI arrays have achieved significant milestones: Hua et al. reported a 64×64 photonic accelerator integrating over 16,000 components with 2.5D hybrid packaging, achieving sub-10 ns latency per multiply-accumulate (MAC) cycle [54]; Lin et al. demonstrated a 120 GOPS photonic tensor core in thin-film lithium niobate (TFLN), enabling both inference and in situ training [55]; and Ahmed et al. presented a quad-core photonic processor integrating four 128×128 MZI-based tensor cores, achieving 65.5 TOPS and near-electronic precision on advanced AI models including ResNet and BERT [56]. These milestones have established MZI arrays as programmable linear-optical processors central to emerging optical AI workloads (see also Refs.[45, 57 – 59]).

**Phase modulation: diffractive structures**

Diffractive architectures are gaining attention for their scalability and energy efficiency. Unlike MZI meshes, where the number of components grows quadratically with the input dimension, diffractive designs require only a linearly scaling number of phase-modulating elements per layer, substantially reducing hardware complexity. In these architectures, each layer consists of locally programmable elements that adjust the refractive index to apply phase shifts; between successive layers, optical diffraction—emulated in integrated implementations via the discrete Fourier transform (DFT)—redistributes the field and reshapes the output intensity distribution (Fig. 3b(i)).

Several integrated implementations have demonstrated on-chip diffractive neural networks, using either passive DFT elements on silicon-on-insulator chips [60] or silicon-slot-based weight encoding [61] (Fig. 3b(ii–iii)). Although initial designs are largely static, reconfigurability has been achieved via thermo-optic tuning [62], optical pumping on active platforms [63], and integrated phase-change materials (PCMs) for nonvolatile weight storage [64].

Beyond single-chip implementations, distributed architectures offer a pathway to larger scales. Xu et al. introduced Taichi, a photonic chiplet achieving 160 TOPS/W energy efficiency through a distributed diffractive architecture [65]. By cascading multiple chiplets, this system demonstrated a network with 13.96 million artificial neurons, overcoming the scalability limits of monolithic integration.

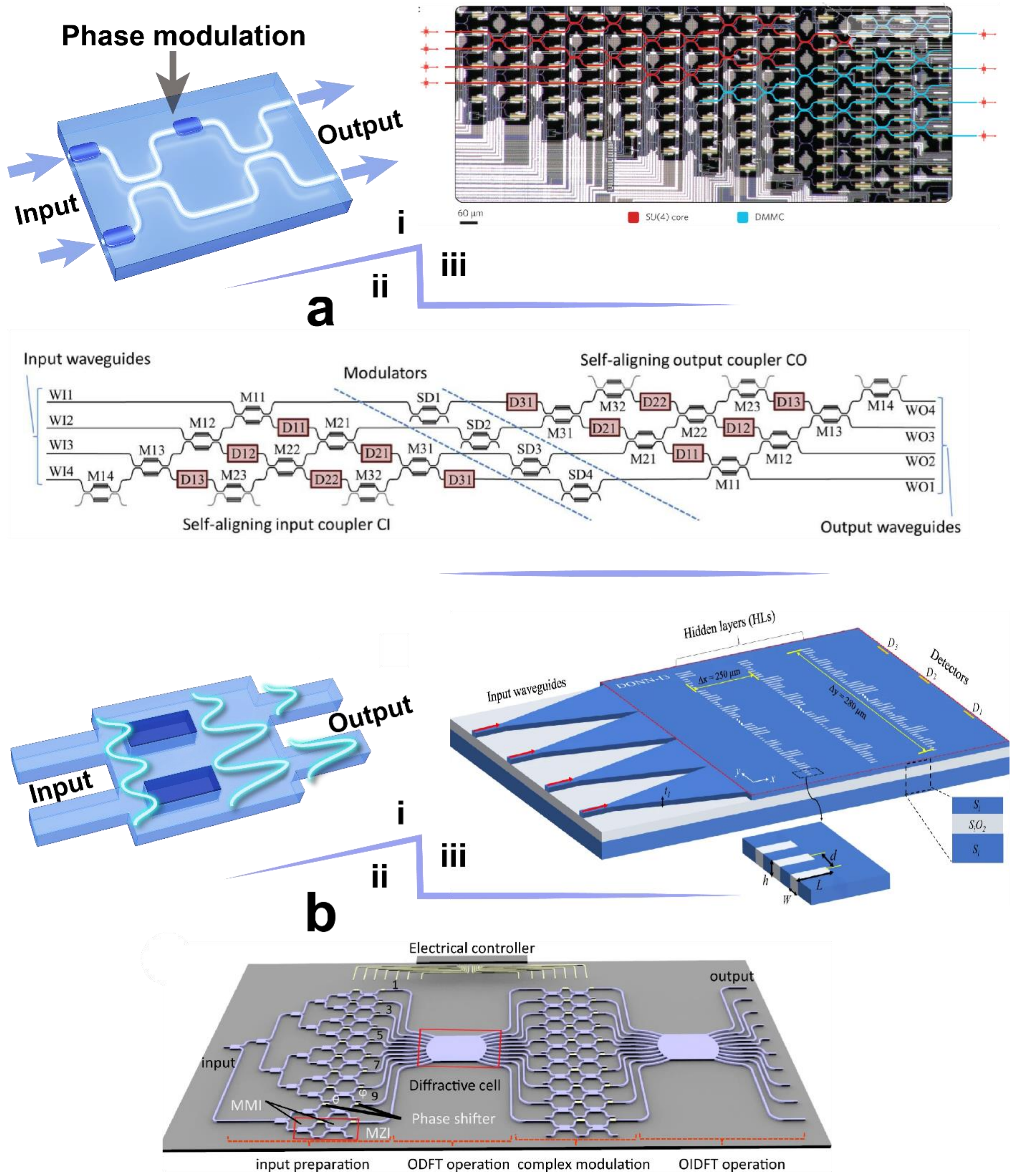

**Fig. 3 Methods of utilising phase modulation for photonic computing in low-dimensional architectures.** a(i) Schematic of a basic unit of MZI for phase-based photonic computing. (ii) An MZI mesh implementing a reconfigurable photonic computing system [50]. (iii) Optical system based on a multilayer MZI structure for vowel recognition. b(i) Schematic of a basic diffractive unit [48]. (ii) Schematic of an integrated neural network with diffractive cells specifically designed for performing DFT and inverse DFT operations [60]. (iii) Schematic of a three-layer diffractive-based implementation of an on-chip ONN [61]. a(ii) Reproduced with permission from ref. [50] from © Chinese Laser Press. a(iii) Reproduced from ref. [48] with permission of Springer Nature: Nature Photonics. b(ii,iii) Reproduced from refs. [60, 61] with permission of Springer Nature: Nature Communications.

### Amplitude modulation: microring resonators

The MRR is a compact device consisting of a closed-loop waveguide that couples evanescently to straight waveguides [66, 67]. When the optical path length matches an integer multiple of the wavelength, resonance produces sharp spectral features—a transmission notch at the through port and a corresponding peak at the drop port (Fig. 4a(i)) [68]. By shifting these resonance peaks—via thermal, electrical, or mechanical perturbations—the transmitted intensity at a fixed wavelength can be continuously varied, enabling precise amplitude modulation. Owing to their tunable resonances, low power consumption, and large free spectral range enabling dense WDM, MRRs have become key building blocks in photonic computing platforms [69, 70].

In a typical silicon photonic accelerator (Fig. 4a(ii)), each MRR is tuned to a specific wavelength and encodes a corresponding weight through thermo-optic [71], electro-optic [72], or mechanical actuation [73]. To realise both positive and negative weights, the system uses through and drop paths routed to balanced photodetectors, whose subtraction yields a signed output proportional to the weight magnitude. The output current is given by:

$$Y_{\mathbf{out}} = \int_{-\infty}^{\infty} |E_0(\lambda)|^2 X_{\mathbf{in}}(\lambda) W_{\mathbf{dt}}(\lambda) R(\lambda) \mathbf{d}\lambda,$$

where $E_0(\lambda)$ is the input optical field, $R(\lambda)$ the detector responsivity, and $W_{\mathbf{dt}}(\lambda)$ the difference in intensity transmission between the drop and through paths. Extending this single-ring principle to matrix operations, an alternative MRR crossbar array approach (Fig. 4a(iii)) computes the transpose matrix operation when WDM light enters from the opposite direction [72, 74].

**Amplitude modulation: absorptive structures**

Absorptive structures implement amplitude-based modulation by dynamically tuning the transmitted optical power. A widely adopted mechanism is evanescent coupling between a waveguide and a PCM, where changes in the PCM phase state alter the coupling strength and, consequently, the transmission amplitude (Fig. 4b(i)) [30, 75–78].

In photonic tensor cores, PCMs directly encode matrix weights through their tunable transmission levels. With transmission contrast exceeding 20%, PCMs enable precise analog weight control and support parallel MVM in photonic crossbar arrays (Fig. 4b(ii)) [79–81].

Beyond PCMs, carrier-tunable semiconductor structures such as P-doped-intrinsic-N-doped (PIN) attenuators offer an alternative, in which forward bias adjusts the absorption coefficient (Fig. 4b(iii)) [82]. Both approaches provide compact, energy-efficient pathways for amplitude-based operations in photonic neural networks.

To conclude, phase and amplitude modulation techniques have reached a high level of maturity and form the backbone of current photonic computing systems. However, each physical channel typically carries a single data stream, so that scaling throughput requires a proportional increase in component count and chip area. The next stage of progress will likely emerge from architectures that exploit additional degrees of freedom beyond phase and amplitude, harnessing the intrinsic parallelism of light to achieve more scalable and efficient photonic computing.

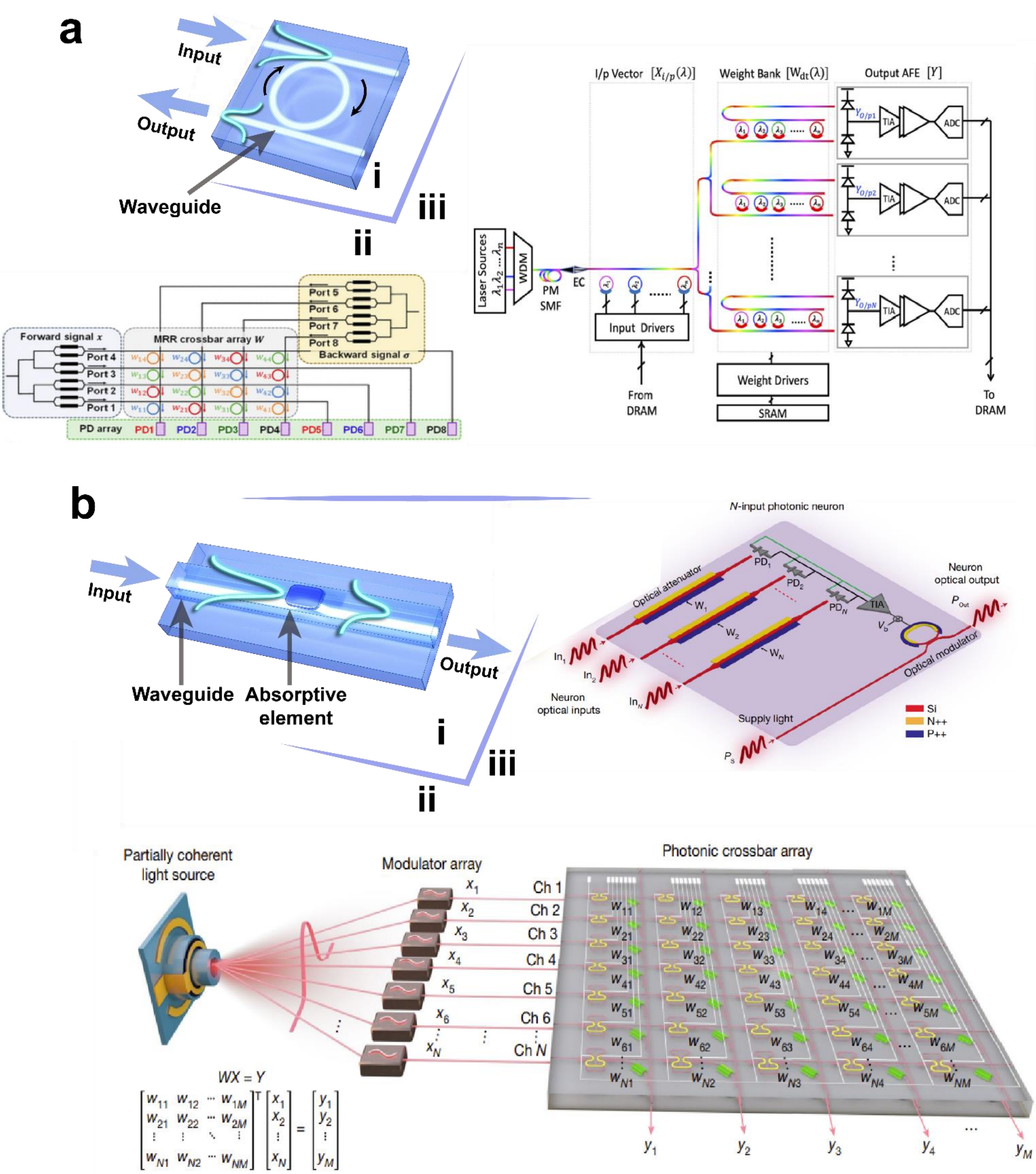

**Fig. 4 Methods of utilising amplitude modulation for photonic computing in low-dimensional architectures.** a(i) Schematic of an MRR for amplitude-based photonic computing. (ii) Circuit diagram of an $N \times N$ accelerator using MRRs in a silicon photonics platform [72]. (iii) An MRR crossbar array for MVM in photonic computing [83]. b(i) Schematic showing the role of PCMs integrated into waveguides for amplitude modulation. (ii) Photonic computing system employing PCMs as the computational core [80]. (iii) Architecture of an N-input optical neuron, where computation is predominantly performed using absorptive elements [82]. a(ii) Reprinted from ref. [72] with the permission of ACS Publications. a(iii) Reproduced from ref. [83] with the permission of AIP Publishing. b(ii) Reproduced from ref. [80] with permission of Springer Nature: Nature. b(iii) Reproduced from ref. [82] with permission of Springer Nature: Nature.

## 3.2 High-dimensional computing

High-dimensional photonic computing exploits spatial modes and spectral channels to enable large-scale parallel processing (Fig. 5a). Crucially, the computational units remain the same as those in low-dimensional computing, ensuring compatibility with established device designs. The key advantage of high-dimensional approaches lies in their ability to multiplex multiple independent data streams through shared hardware, thereby increasing throughput without proportionally scaling the number of active components.

**Spatial modes**

The spatial-mode approach encodes information into the transverse electric field profile, enabling parallel processing through MDM. As shown in Fig. 5b, MDM utilises distinct optical modes, ranging from $\mathbf{TE}_0$ to $\mathbf{TE}_3$, within a waveguide to increase channel capacity and support simultaneous data streams [84–86]. A typical MDM-based system integrates mode multiplexers, multimode beam splitters, modulator arrays, and signal combiners to perform optical MVM [87].

In the architecture shown in Fig. 5b, four incoherent mode signals $(\mathbf{TE}_0 - \mathbf{TE}_3)$ are multiplexed, processed through a 4×4 MZI modulator matrix, and recombined to yield the weighted sum. The output vector is given by:

$$O = MI,$$

where $M$ is the modulation matrix and $I$ is the optical input vector. By carrying multiple data channels within a single physical waveguide, MDM multiplies the effective throughput by the number of supported modes—a four-mode system, for instance, achieves a fourfold increase in channel capacity relative to single-mode operation [87]. This scaling comes without requiring additional wavelengths or separate waveguides, making MDM particularly attractive for area-constrained on-chip designs.

However, inter-mode crosstalk—arising from fabrication imperfections, waveguide bends, and imperfect mode (de)multiplexing—limits the number of modes that can be reliably used in practice [85,86]. Current on-chip demonstrations typically support up to four to eight modes, and the insertion loss of mode converters grows with the mode count, imposing a practical trade-off between parallelism and signal fidelity [84]. Nonetheless, ongoing advances in waveguide design, fabrication precision, and crosstalk-mitigation algorithms are steadily pushing these limits, suggesting that MDM-based photonic computing has considerable room for further scaling.

**Spectral multiplexing**

A complementary high-dimensional route exploits spectral multiplexing, most commonly through WDM, to achieve parallelism [88–94]. As shown in Fig. 5c, an optical frequency comb provides multiple carriers that are typically addressed as wavelength channels, each interacting with weight cells to implement real-time MVM. A representative implementation is the photonic tensor core demonstrated by Feldmann et al. [30], in which a soliton microcomb provides multiple wavelength channels that each pass through PCM-based weight cells on a 16×16 matrix; the weighted optical signals are summed on a photodetector to realise a dot product. Because all wavelength channels operate simultaneously, the system achieves a throughput of 2 tera-MAC operations per second at a modulation bandwidth exceeding 14 GHz, with an optical energy cost of only 17 fJ per MAC operation. The authors project that, with moderate scaling to a 40×40 matrix using foundry-standard silicon-on-insulator processes, compute densities exceeding 400 TOPS mm$^{-2}$ are attainable [30]. These figures concretely demonstrate how WDM translates spectral bandwidth into computational throughput.

WDM benefits from decades of development in optical telecommunications, with mature components—arrayed waveguide gratings, micro-ring filters, and erbium-doped amplifiers—readily available for integration [90,91]. Compared with MDM, WDM offers a larger number of usable channels (tens to over one hundred wavelengths from a single comb source), providing greater parallelism per waveguide. The principal challenges include the thermal sensitivity of resonance-based filters, which require active stabilisation to maintain channel alignment, and the finite free spectral range of MRRs, which bounds the maximum channel count for a given spectral window [91].

A further route to improving WDM efficiency is to revisit the coherence requirements of the light source itself. Dong et al. [80] demonstrated that using partially coherent light allows the same wavelength band to be distributed across all $N$ input channels of an $N \times N$ tensor core without coherent interference artefacts, whereas a fully coherent system would require $N$ distinct wavelengths—yielding an $N$-fold improvement in spectral efficiency. Using a 9×3 silicon photonic tensor core with integrated electro-absorption modulators, the system achieves 0.108 TOPS at a data rate of 2 GSa/s per channel, with an estimated energy efficiency of approximately 1 TOPS W$^{-1}$ [80]. These results suggest that relaxing coherence requirements—rather than tightening them—can be a viable route to scaling photonic tensor cores. Beyond improving individual domains, advanced implementations further expand the dimensionality by combining spatial, spectral, and radio-frequency domains in a single chip. Such hybrid approaches provide a foundation for future systems that integrate multiple encoding domains within a unified photonic platform.

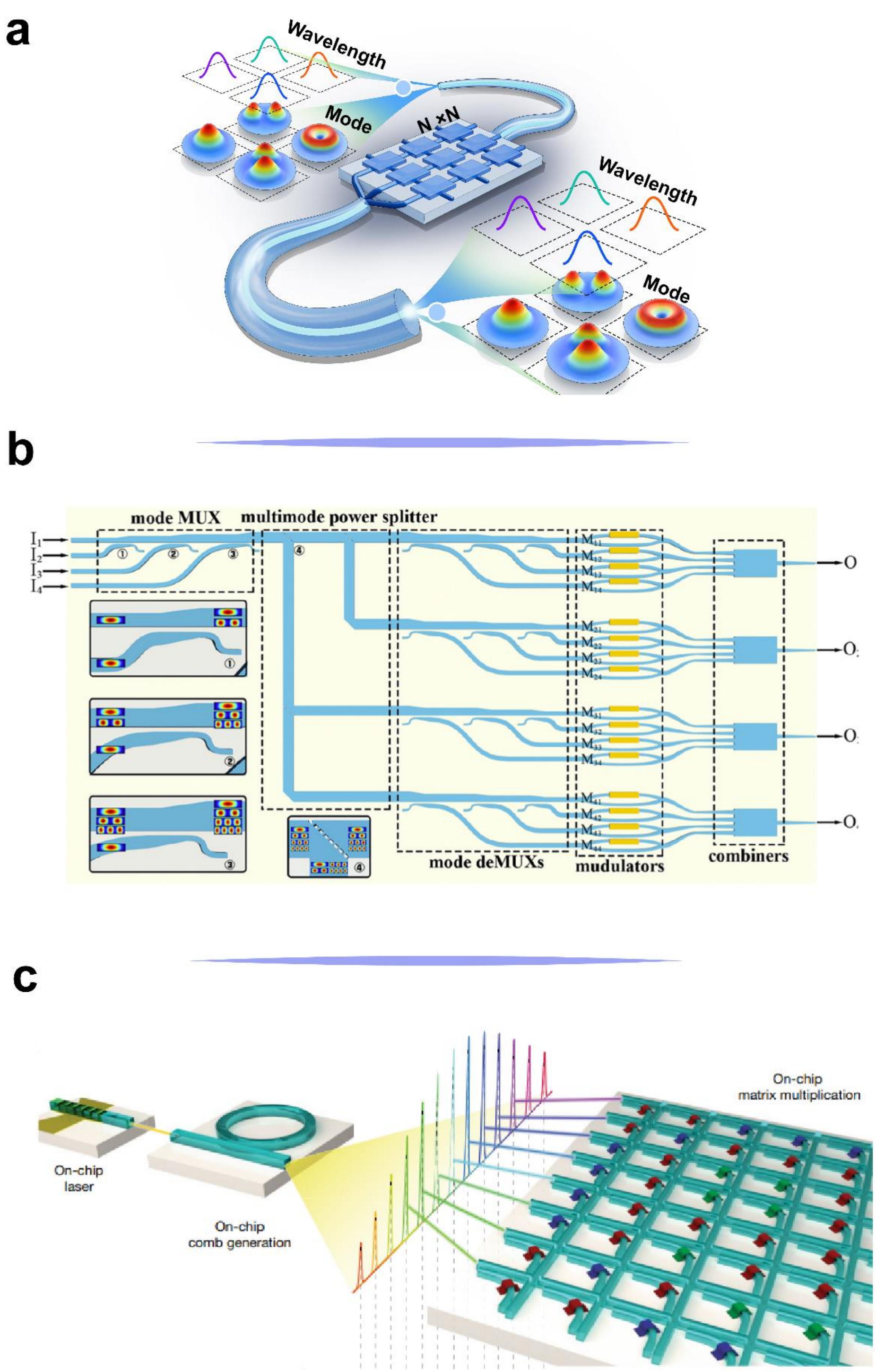

**Fig. 5 High-dimensional photonic computing for efficient data transmission.** (a) Schematic of high-dimensional transmission, where multiple spatial modes or spectral channels are transmitted through a single waveguide. (b) System for multi-mode processing, where multiple modes are combined, demultiplexed for independent computation, and recombined for detection [87]. (c) Schematic of an optical convolutional neural network architecture, illustrating the use of wavelength-multiplexed inputs generated by a frequency comb. Each wavelength channel encodes an input element that interacts with an array of PCM cells, enabling convolutional MVMs within the photonic core [30]. (b) Reproduced from ref. [87] with the permission of Elsevier. (c) Reproduced from ref. [30] with permission of Springer Nature: Nature.

In summary, high-dimensional approaches have demonstrated concrete performance gains: WDM-based tensor cores already achieve tera-MAC-scale throughput ($\sim 2 \times 10^{12}$ MAC $s^{-1}$ [30]), while MDM multiplies channel capacity by the number of supported modes, and partial-coherence techniques further improve spectral efficiency by up to $N$-fold [80]. The two principal multiplexing strategies—MDM and WDM—offer complementary trade-offs: MDM provides moderate parallelism (four to eight channels) with compact, wavelength-agnostic designs, whereas WDM supports larger channel counts but requires thermal stabilisation and broadband comb sources. Beyond these established approaches, emerging topological structures such as optical skyrmions offer additional encoding dimensions by exploiting discretised topological invariants that are inherently robust against perturbations (see Section 4 for further discussion). Realising the full potential of high-dimensional computing, however, requires not only careful device integration but also system-level coordination. The following section examines such strategies.

### 3.3 System-level optimisations

Beyond device-level modulation, system-level strategies play an equally important role in realising practical photonic computing. While individual photonic architectures differ in their physical implementations, they share a common set of bottlenecks: limited electronic-to-optical conversion bandwidth, high per-node hardware cost, restricted numerical representation, and sensitivity to fabrication and thermal non-idealities. This section examines five strategies, each addressing one of these cross-cutting challenges, and illustrates each with a representative implementation. Importantly, the underlying principles—reducing interface overhead, amortising resources, expanding representable weight spaces, and co-designing training with hardware constraints—are transferable across the MZI, MRR, diffractive, and PCM-based architectures discussed in Sections 3.1 and 3.2.

**Time and wavelength interleaving**

In photonic computing systems, the interface between the digital electronics and photonic hardware often determines overall performance. A key bottleneck in photonic computing lies in the digital-to-analog converters (DACs) that drive optical modulators: a single high-speed DAC must satisfy demanding requirements in both sampling rate and bit resolution, and its power consumption grows super-linearly with these specifications. To relax these requirements, Xu et al. [95] introduced a time-wavelength interleaving strategy (Fig. 6a). Rather than feeding the entire input vector through one modulator at full speed, the data stream is partitioned across $N$ wavelength channels and $T$ consecutive time slots. Each individual DAC-modulator pair therefore operates at only $1/(N \times T)$ of the aggregate data rate, substantially lowering the per-channel speed and resolution requirements while preserving the total system throughput—which reached 11.32 TOPS in their demonstration [95].

This joint encoding strategy offers two concrete advantages. First, by reducing the instantaneous symbol rate per channel, it enables the use of lower-speed, lower-power DACs without sacrificing aggregate throughput. Second, the broader distribution of data across spectral and temporal domains improves energy efficiency: the power required per conversion decreases, and the broad optical bandwidth of the comb source is more fully utilised. Together, these benefits make time-wavelength interleaving an effective system-level approach to overcoming the electronic-to-optical conversion bottleneck.

**Amortisation**

While the preceding strategies target on-chip architectures, photonic computing can also benefit from system-level amortisation in distributed settings, where the costs of weight generation, optical accumulation, and electronic readout are shared across many operations or users. Sludds et al. [96] demonstrated a server—client architecture (Fig. 6b) in which a central server broadcasts neural network weights to lightweight edge devices using amplitude-modulated optical fields across multiple wavelengths. Each client, equipped with only a single optical modulator, performs MVM locally, while a time-integrating receiver accumulates the weighted signals before a single electronic readout.

The key idea is to amortise the cost of weight generation and distribution across many edge nodes, so that each client avoids the need for on-chip weight storage, large-scale analog-to-digital converter (ADC) arrays, or active photonic components. This amortisation enables an energy efficiency of 0.1 aJ per MAC—equivalent to fewer than one photon per operation [96, 97]—because the computation is accumulated in the optical domain before conversion. Although this architecture is specific to distributed inference scenarios rather than general-purpose on-chip accelerators, it illustrates a broader principle [98, 99]: system-level design choices about where weights are generated, stored, distributed, accumulated, and read out can yield efficiency gains complementary to device-level improvements.

**Coherent detection**

A third cross-cutting bottleneck concerns numerical representation. In any photonic architecture using incoherent (intensity-only) detection, the measurable quantities are inherently real-valued and non-negative, restricting the weight space to positive reals and limiting computational expressiveness. This constraint affects MZI meshes, MRR weight banks, and PCM-based tensor cores alike, because all rely on photodetectors that discard phase information. Coherent detection overcomes this limitation by preserving the optical phase through homodyne or heterodyne readout, thereby enabling native support for negative and complex-valued weights (Fig. 6c(i)). As an illustrative implementation, Rahimi Kari et al. [100] demonstrated a coherent dot-product architecture using eight-channel current sources to modulate both phase and amplitude of input signals (Fig. 6c(ii)). Thermo-optic MZIs serve as beamsplitters and phase shifters for complex-valued encoding, and homodyne detection captures both quadratures without additional post-processing. The general principle—phase-sensitive readout to expand the representable weight space—is applicable to any photonic computing platform and eliminates the need for high-resolution ADCs that would otherwise be required to encode sign information electronically.

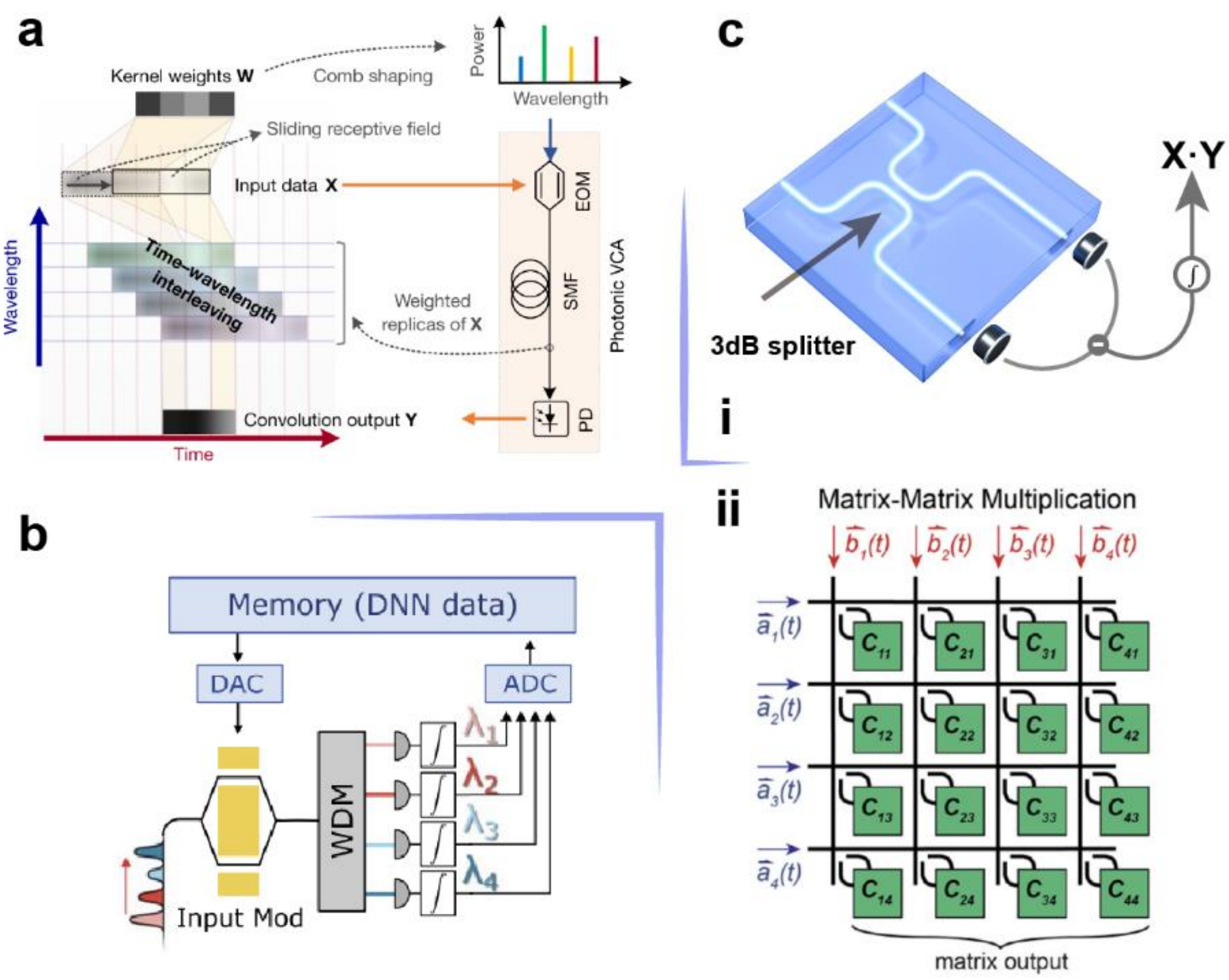

**Fig. 6 Representative system-level strategies for improving efficiency in photonic computing.** (a) Schematic of a time-wavelength interleaving scheme for parallel data processing [95]. (b) Schematic of an integration circuit at the receiver node [96]. (c)(i) Schematic of an integrated dot-product unit cell. (ii) Schematic of a two-dimensional array of dot-product unit cells for performing matrix-matrix multiplication, with rows and columns supplied by time-multiplexed input signals [100]. (a) Reproduced from ref. [95] with permission of Springer Nature: Nature. (b) Reproduced by permission from AAAS [96]. (c) Reproduced with permission from ref. [100] from © Optica Publishing Group.

### Efficient learning methods

Traditional neural network training relies on backpropagation, which requires accurate modelling of each layer and involves complex gradient computations [101]. To reduce reliance on external computational resources, two alternative learning methods have been developed for ONNs: fully forward mode (FFM) and in situ backpropagation.

FFM eliminates backward signal propagation by using two forward passes to estimate gradients, leveraging optical reciprocity and spatial symmetry [102]. In situ backpropagation encodes error signals optically and propagates them backward through the network, extracting gradients from optical interference between error and inference signals [48, 103, 104]. Both approaches shift gradient estimation into the optical domain, offering pathways towards energy-efficient and scalable training for ONNs.

### Hardware-aware training

ONNs, particularly those based on MRRs, are highly sensitive to thermal and fabrication noise [105]. Hardware-aware training improves both energy efficiency and reliability by tailoring the learning process to the nonlinear characteristics of optical components.

In this approach, neural networks are first trained conventionally and then fine-tuned with a regularisation term that shifts the weights away from thermally unstable regions, typically near 0 [105], towards more stable operating points, typically near 1. This selective regularisation enhances system resilience to resonance drift without sacrificing accuracy [105]. The training process, therefore, consists of two stages: a standard optimisation phase followed by hardware-aware fine-tuning. The outcome is a more robust and power-efficient system that maintains high performance in realistic optical environments while reducing the need for additional control circuitry.

The device-level and system-level strategies discussed in this section-have collectively advanced photonic computing from laboratory demonstrations toward practical deployment. However, translating these advances into scalable, real-world systems requires addressing several fundamental challenges that span materials, architectures, and algorithms. The following section examines these challenges and outlines promising directions for future research.

## 4. Challenges and perspectives

Enhancing the computational capabilities of optical chips hinges on scaling matrix sizes and improving parallelism [80, 106]. Larger matrices help amortise the fixed energy cost of electro-optic conversions, while high-dimensional techniques further boost throughput (see Supplementary Information for detailed analysis).

Scaling up, however, also introduces new challenges. As outlined in Section 1, we examine five major areas that currently limit scalability and practicality: electro-optic efficiency, computing parallelism, spatial integration, reconfigurability, and robustness.

### Electro-optic efficiency

Despite their inherent advantages in speed and bandwidth, optical systems remain constrained by the energy costs of electro-optic conversion (Fig. 7a). In practice, this is dominated by ADCs and DACs, which account for a large fraction of system-level power consumption and offset much of the energy advantage offered by optical data processing [99]. As detailed in the Supplementary Information, this bottleneck represents a primary hurdle to achieving ultra-low energy per operation in current optical architectures.

Increasing matrix sizes can help amortise the static energy cost of conversion, but sustained improvements require targeted innovations. These may include significantly improving ADC and DAC efficiency or developing fully optical systems that avoid conversion altogether [107]. Lowering the power consumption of peripheral electronic components, such as interconnects and drivers, is also important for narrowing the energy gap between optical and electronic systems [108, 109]. Continued progress in electro-optic interface optimisation will therefore be critical for advancing scalable and energy-efficient photonic computing [110, 111].

Beyond energy efficiency at the conversion interface, another key avenue for improving system performance lies in expanding the computational throughput through parallelism.

### Computing parallelism

While phase/amplitude modulation and MDM/WDM-based multiplexing have become established routes in photonic computing, other degrees of freedom remain comparatively underutilised (Fig. 7b). Uniform polarisation states represent one underexplored avenue. On-chip polarisation control has been demonstrated using birefringent waveguides and anisotropic materials, and polarisation-modulating components are increasingly available [112 - 114]. However, polarisation offers only two mutually orthogonal states (e.g., horizontal/vertical or left/right circular), fundamentally capping the channel capacity at a factor of two—far less than the tens to hundreds of channels achievable with WDM or MDM.

Another spatial-mode strategy employs orbital angular momentum (OAM) modes as an alternative basis for MDM. Because OAM modes are themselves spatial eigenmodes, expressible as superpositions of conventional guided-mode profiles ($TE_0$, $TE_1$, ...), OAM-based multiplexing is fundamentally a form of MDM rather than a distinct physical mechanism. Recent demonstrations of OAM mode transmission through integrated waveguides [115, 116] confirm the on-chip viability of this approach. A qualitatively different direction is topological encoding, which exploits integer-valued invariants, such as skyrmion number, that are inherently robust against continuous perturbations. Unlike conventional mode indices, topological invariants are discretised and resistant to fabrication imperfections and thermal drift, offering intrinsic fault tolerance. Skyrmion-based logic has been implemented in nanophotonic platforms [117, 118], and meta-fiber experiments have further demonstrated the transmission of structured polarisation fields, suggesting that topological architectures could be adapted for on-chip photonic computing systems [119].

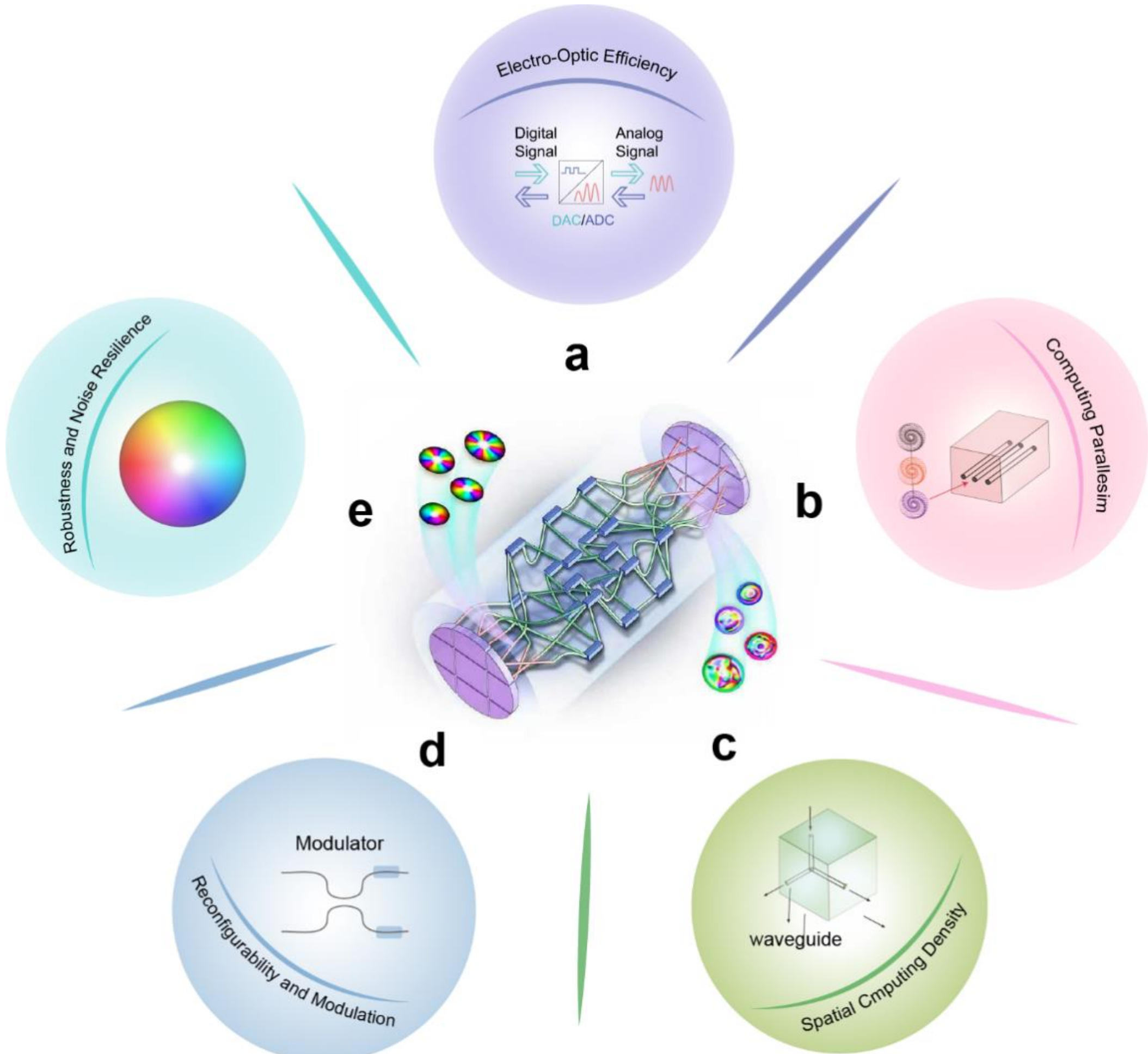

**Fig. 7 Future opportunities for photonic computing.** (a) Electro-optic conversion efficiency, highlighting hardware and architectural strategies for improving energy efficiency. (b) Computing parallelism, illustrating the use of other degrees of freedom of light to increase information density and efficiency. (c) Spatial computing density, illustrating the use of alternative technologies such as direct laser writing to create three-dimensional structures beyond planar geometries. (d) Reconfigurability and modulation, highlighting material and design advances to improve the integration density, speed, and device flexibility. (e) Robustness and noise resilience, showing algorithmic and design improvements to strengthen photonic computing systems against noise and error.

Although these approaches are still at an early stage, integrating uniform polarisation control and spatially structured fields with existing phase/amplitude modulation and MDM/WDM-based methods offers the potential to unlock new forms of parallelism. Such advances may provide the foundation for massively parallel, high-dimensional photonic computing architectures capable of handling increasingly demanding workloads.

While exploiting additional degrees of freedom can enhance parallelism, the physical layout of optical circuits also imposes fundamental constraints on scalability. Addressing these spatial limitations requires moving beyond conventional planar architectures.

**Spatial integration and 3D scaling**

Most integrated optical systems are currently built on planar silicon platforms using traditional semiconductor fabrication processes (Fig. 7c). While this technology is relatively mature, it is fundamentally constrained in spatial density by its two-dimensional layout. The absence of vertical interconnects not only limits the number of optical elements that can be integrated on-chip but also forces nonadjacent components to connect through planar waveguides, introducing additional loss at intersections [120]. These limitations represent a significant obstacle to scaling optical circuits for large-scale applications.

Three-dimensional integration offers a path to overcome these constraints by enabling greater design flexibility and improved utilisation of light. Direct laser writing, for example, allows the fabrication of sub-micron-scale structures in glass or polymer through nonlinear processes. Recent demonstrations of 3D interconnects inside optical chips show the potential for vertical stacking and cross-layer routing, which could reduce circuit footprint while improving scalability [121–123]. Moreover, additively printed components have been successfully integrated with semiconductor samples, indicating growing CMOS compatibility and opening opportunities for compact packaging of 3D optical devices [124].

Additional advances are being made with vertical-cavity surface-emitting lasers (VCSELs), which emit light out of plane and thus naturally lend themselves to 3D architectures [125]. VCSELs are also CMOS-compatible and exhibit nonlinear responses to both optical and electronic signals, making them promising candidates for compact optical neurons in vertically integrated networks.

Finally, industry initiatives are accelerating progress towards 3D scaling. Companies such as TSMC are actively developing multilayer photonics, leveraging expertise in silicon and silicon nitride (SiN) devices to enable heterogeneous integration of optical and electronic circuits through chip-on-wafer-on-substrate packaging [126]. Advances in CMOS fabrication technology are also inspiring new approaches to silicon photonics, potentially allowing 3D architectures to be realised directly at the fabrication level [127, 128]. Together, these efforts highlight the momentum towards spatially integrated and vertically scalable photonic computing platforms.

As integration density increases, the ability to dynamically program and precisely control optical weights becomes increasingly important for realising flexible and adaptive photonic computing systems.

**Reconfigurability**

The ability to dynamically configure and precisely control optical weights is central to photonic computing, particularly for MVM operations underlying machine learning tasks, while system-level control of signal pathways becomes increasingly important as architectures scale (Fig. 7d). Several modulation mechanisms have been developed, with thermo-optic and electro-optic approaches currently dominating. Thermo-optic modulators exploit silicon's temperature-dependent refractive index but suffer from high power consumption, slow response, and thermal crosstalk, limiting device density and scalability [129]. Electro-optic modulators, especially hybrid lithium niobate ($LiNbO_3$) platforms exploiting the Pockels effect, offer significantly faster response times but face footprint constraints [130]. Alternative approaches, including liquid-crystal [131-133], micro-electro-mechanical systems (MEMS) [134], and acousto-optic modulators [73], are also being explored, each with distinct trade-offs between speed, power, and integration density. At the system level, coarse-grained optical switching and reconfigurable interconnects may further support resource sharing, topology adaptation, and fault bypass among photonic cores, provided that insertion loss, crosstalk, and control overheads are carefully co-optimised.

Moreover, precision in modulation remains a critical challenge for scaling. Small errors in weight settings can propagate and accumulate across large circuits, degrading computational accuracy. Addressing this may require improvements in device uniformity, real-time calibration schemes, and error-tolerant algorithms. Together, advances in both modulation technologies and precision control will be essential for realising scalable, high-performance photonic computing systems.

Even with precise modulation and reconfigurable architectures, photonic systems must contend with inherent noise and hardware imperfections that can compromise computational fidelity at scale.

**Robustness and noise resilience**

ONNs, due to their analog nature, are more vulnerable to noise and hardware variation than digital electronic systems (Fig. 7e). Common sources of error include phase drift in interferometric circuits, thermal crosstalk in resonator arrays, and fabrication imperfections. As systems scale, these effects accumulate and degrade computational accuracy [105].

Noise-aware training offers one solution by incorporating noise models directly into the optimisation process, enhancing error tolerance without substantial hardware overhead [135].

Beyond algorithmic approaches, emerging topological strategies offer a fundamentally different route to robust information encoding. Optical skyrmions [119] exploit discretised topological invariants that are inherently resilient to perturbations such as thermal drift and fabrication imperfections [136]. Unlike conventional phase- or amplitude-based encoding, skyrmion-based platforms enable digitisation of analog optical fields through integer-valued topological numbers, offering intrinsic fault tolerance for logic operations and information storage [117, 118]. Because the encoding relies on a topological invariant rather than a continuously tunable parameter, it is qualitatively more robust than the waveguide-mode channels used in MDM (Section 3.2). These advances suggest that topological encoding may provide a foundation for scalable, noise-resilient photonic computing systems.

## 5. Conclusion

The rapid growth of AI has intensified demands on computing infrastructure, yet conventional CMOS electronics face fundamental limits in bandwidth, energy efficiency, and parallelism. These constraints have motivated the search for alternative computing paradigms capable of sustaining next-generation AI workloads.

This review has surveyed the foundations and recent advances of integrated photonic computing as a scalable and energy-efficient alternative. At its core, photonic computing exploits light's intrinsic physical degrees of freedom to enable parallel, high-throughput computation. We place particular emphasis on high-dimensional computing architectures that leverage the full dimensionality of light. Our discussion is organised around a two-level progression:

(1) Low-dimensional computing directly modulates phase and amplitude using compact optical components such as MZIs, diffractive structures, MRRs, and absorptive elements, providing a high-speed foundation for optical signal processing.

(2) High-dimensional computing extends these capabilities by leveraging additional degrees of freedom, including spatial-mode multiplexing, spectral multiplexing, and emerging topological structures such as optical skyrmions, to dramatically scale computational density with only modest increases in hardware complexity.

Beyond device-level advances, the five system-level strategies examined in Section 3.3 address bottlenecks in energy efficiency and precision, highlighting the growing importance of algorithm-hardware co-design.

Despite these advances, several challenges remain. ADC/DAC operations continue to dominate system-level energy budgets. Hardware non-idealities, including thermal drift and fabrication variation, constrain scalability. Moreover, many degrees of freedom of light, particularly polarisation and topological photonic states such as optical skyrmions, remain underutilised but offer promising opportunities for fault-tolerant, high-dimensional encoding. Meeting these challenges will require coordinated efforts across materials science, device engineering, and system architecture. We believe that, by embracing the higher dimensionality of light, including emerging topological approaches such as optical skyrmions, photonic computing offers not merely an evolutionary improvement but a fundamentally new paradigm for high-performance and sustainable information processing.

**References:**

1. Ning, S. *et al.* Photonic-electronic integrated circuits for high-performance computing and AI accelerators. *J. Light. Technol.* **42**, 7834–7859 (2024).
2. Lundstrom, M. Moore's Law forever? *Science* **299**, 210–211 (2003).
3. Baker, R. J. *CMOS Circuit Design, Layout, and Simulation*. (IEEE Press, Piscataway, NJ, 2019).
4. Radamson, H. H. *et al.* State of the art and future perspectives in advanced CMOS technology. *Nanomaterials* **10**, 1555 (2020).
5. Horowitz, M. *et al.* Scaling, power, and the future of CMOS. in *IEEE International Electron Devices Meeting, 2005. IEDM Technical Digest.* 9–15 (IEEE, Tempe, Arizona, USA, 2005). doi:10.1109/IEDM.2005.1609253.
6. Zhao, Y., Gobbi, M., Hueso, L. E. & Samorì, P. Molecular approach to engineer two-dimensional devices for CMOS and beyond-CMOS applications. *Chem. Rev.* **122**, 50–131 (2022).
7. Bespalov, V. A., Dyuzhev, N. A. & Kireev, V. Yu. Possibilities and limitations of CMOS technology for the production of various microelectronic systems and devices. *Nanobiotechnology Rep.* **17**, 24–38 (2022).
8. Krambeck, R. H., Lee, C. M. & Law, H.-F. S. High-speed compact circuits with CMOS. *IEEE J. Solid-State Circuits* **17**, 614–619 (1982).
9. Hill, I., Chanawala, P., Singh, R., Sheikholeslam, S. A. & Ivanov, A. CMOS reliability from past to future: a survey of requirements, trends, and prediction methods. *IEEE Trans. Device Mater. Reliab.* **22**, 1–18 (2022).
10. Teng, Q. *et al.* Reliability challenges in CMOS technology: a manufacturing process perspective. *Microelectron. Eng.* **281**, 112086 (2023).
11. Leiserson, C. E. *et al.* There's plenty of room at the top: What will drive computer performance after Moore's Law? *Science* **368**, eaam9744 (2020).
12. Shalf, J. The future of computing beyond Moore's Law. *Philos. Trans. R. Soc. Math. Phys. Eng. Sci.* **378**, 20190061 (2020).
13. Lattner, C. *et al.* MLIR: a compiler infrastructure for the end of Moore's Law. Preprint at https://doi.org/10.48550/arXiv.2002.11054 (2020).
14. Johnsson, L. & Netzer, G. The impact of Moore's Law and loss of Dennard scaling: Are DSP SoCs an energy efficient alternative to x86 SoCs? *J. Phys. Conf. Ser.* **762**, 012022 (2016).
15. Huang, Y., Guo, B. & Shen, Y. GPU energy consumption optimization with a global-based neural network method. *IEEE Access* **7**, 64303–64314 (2019).
16. Mittal, S. & Vetter, J. S. A survey of methods for analyzing and improving GPU energy efficiency. *ACM Comput. Surv.* **47**, 1–23 (2015).
17. You, J., Chung, J.-W. & Chowdhury, M. Zeus: Understanding and optimizing GPU energy consumption of DNN training. in *20th USENIX Symposium on Networked Systems Design and Implementation (NSDI 23)* 119–139 (USENIX Association, Boston, MA, 2023).
18. Patterson, D. *et al.* Carbon emissions and large neural network training. Preprint at https://doi.org/10.48550/arXiv.2104.10350 (2021).
19. Brown, T. B. *et al.* Language models are few-shot learners. Preprint at https://doi.org/10.48550/arXiv.2005.14165 (2020).
20. Jot Singh, J., Dhawan, D. & Gupta, N. All-optical photonic crystal logic gates for optical computing: an extensive review. *Opt. Eng.* **59**, 110901 (2020).
21. Kibebe, C. G., Liu, Y. & Tang, J. Harnessing optical advantages in computing: a review of current and future trends. *Front. Phys.* **12**, 1379051 (2024).
22. McMahon, P. L. The physics of optical computing. *Nat. Rev. Phys.* **5**, 717–734 (2023).
23. Rysin, A., Livshits, P., Sofer, S. & Fefer, Y. Impact of increased resistive losses of metal interconnects upon ULSI devices reliability and functionality. *Microelectron. Eng.* **92**, 119–122 (2012).
24. Al Aziz, M. *et al.* An overview of achieving energy efficiency in on-chip networks. *Int. J. Commun. Netw. Distrib. Syst.* **5**, 444–458 (2010).
25. Burger, D., Goodman, J. R. & Kägi, A. Memory bandwidth limitations of future microprocessors. in *Proceedings of the 23rd annual international symposium on Computer architecture* 78–89 (ACM, Philadelphia Pennsylvania USA, 1996). doi:10.1145/232973.232983.

26. Zou, P. *et al.* Optical computing accelerators: Principle, application, and perspective. *Front. Phys.* **20**, 032302 (2025).
27. Li, R. *et al.* Photonics for neuromorphic computing: Fundamentals, devices, and opportunities. *Adv. Mater.* **37**, 2312825 (2025).
28. Li, C., Zhang, X., Li, J., Fang, T. & Dong, X. The challenges of modern computing and new opportunities for optics. *PhotoniX* **2**, 20 (2021).
29. Pintus, P. *et al.* Integrated non-reciprocal magneto-optics with ultra-high endurance for photonic in-memory computing. *Nat. Photonics* **19**, 54–62 (2025).
30. Feldmann, J. *et al.* Parallel convolutional processing using an integrated photonic tensor core. *Nature* **589**, 52–58 (2021).
31. Aghaee Rad, H. *et al.* Scaling and networking a modular photonic quantum computer. *Nature* **638**, 912–919 (2025).
32. Feng, C. *et al.* Wavelength-division-multiplexing-based electronic-photonic network for high-speed computing (conference presentation). in *Smart Photonic and Optoelectronic Integrated Circuits XXII* (eds He, S. & Vivien, L.) 15 (SPIE, San Francisco, United States, 2020). doi:10.1117/12.2551323.
33. Sotirova, A. S. *et al.* Low cross-talk optical addressing of trapped-ion qubits using a novel integrated photonic chip. *Light Sci. Appl.* **13**, 199 (2024).
34. Zuo, Y. *et al.* All-optical neural network with nonlinear activation functions. *Optica* **6**, 1132 (2019).
35. Sui, X., Wu, Q., Liu, J., Chen, Q. & Gu, G. A review of optical neural networks. *IEEE Access* **8**, 70773–70783 (2020).
36. Zhang, Y. *et al.* Memory-less scattering imaging with ultrafast convolutional optical neural networks. *Sci. Adv.* **10**, eadn2205 (2024).
37. Li, G. H. Y. *et al.* All-optical ultrafast ReLU function for energy-efficient nanophotonic deep learning. *Nanophotonics* **12**, 847–855 (2023).
38. Ho, R., Mai, K. W. & Horowitz, M. A. The future of wires. *Proc. IEEE* **89**, 490–504 (2001).
39. Corcoran, B. *et al.* Ultra-dense optical data transmission over standard fibre with a single chip source. *Nat. Commun.* **11**, 2568 (2020).
40. Desiatov, B., Shams-Ansari, A., Zhang, M., Wang, C. & Lončar, M. Ultra-low-loss integrated visible photonics using thin-film lithium niobate. *Optica* **6**, 380 (2019).
41. Miller, D. A. B. Attojoule optoelectronics for low-energy information processing and communications. *J. Light. Technol.* **35**, 346–396 (2017).
42. Zuo, C. & Chen, Q. Exploiting optical degrees of freedom for information multiplexing in diffractive neural networks. *Light Sci. Appl.* **11**, 208 (2022).
43. Li, J., Hung, Y.-C., Kulce, O., Mengu, D. & Ozcan, A. Polarization multiplexed diffractive computing: all-optical implementation of a group of linear transformations through a polarization-encoded diffractive network. *Light Sci. Appl.* **11**, 153 (2022).
44. Xiang, S. *et al.* A review: Photonics devices, architectures, and algorithms for optical neural computing. *J. Semicond.* **42**, 023105 (2021).
45. Wetzstein, G. *et al.* Inference in artificial intelligence with deep optics and photonics. *Nature* **588**, 39–47 (2020).
46. Bogaerts, W. *et al.* Programmable photonic circuits. *Nature* **586**, 207–216 (2020).
47. Du, Y. *et al.* Implementation of optical neural network based on Mach–Zehnder interferometer array. *IET Optoelectron.* **17**, 1–11 (2023).
48. Shen, Y. *et al.* Deep learning with coherent nanophotonic circuits. *Nat. Photonics* **11**, 441–446 (2017).
49. Harris, N. C. *et al.* Linear programmable nanophotonic processors. *Optica* **5**, 1623 (2018).
50. Miller, D. A. B. Self-configuring universal linear optical component [invited]. *Photon. Res.* **1**, 1 (2013).
51. Miller, D. A. B. Self-aligning universal beam coupler. *Opt. Express* **21**, 6360 (2013).
52. Shokraneh, F., Geoffroy-Gagnon, S. & Liboiron-Ladouceur, O. The diamond mesh, a phase-error- and loss-tolerant field-programmable MZI-based optical processor for optical neural networks. *Opt. Express* **28**, 23495 (2020).
53. Zhou, H. *et al.* Photonic matrix multiplication lights up photonic accelerator and beyond. *Light Sci. Appl.* **11**, 30 (2022).

54. Hua, S. *et al.* An integrated large-scale photonic accelerator with ultralow latency. *Nature* **640**, 361–367 (2025).
55. Lin, Z. *et al.* 120 GOPS photonic tensor core in thin-film lithium niobate for inference and in situ training. *Nat. Commun.* **15**, 9081 (2024).
56. Ahmed, S. R. *et al.* Universal photonic artificial intelligence acceleration. *Nature* **640**, 368–374 (2025).
57. Cheng, J., Zhou, H. & Dong, J. Photonic matrix computing: From fundamentals to applications. *Nanomaterials* **11**, 1683 (2021).
58. Guo, X., Xiang, J., Zhang, Y. & Su, Y. Integrated neuromorphic photonics: Synapses, neurons, and neural networks. *Adv. Photonics Res.* **2**, 2000212 (2021).
59. Tian, Y. *et al.* Scalable and compact photonic neural chip with low learning-capability-loss. *Nanophotonics* **11**, 329–344 (2022).
60. Zhu, H. H. *et al.* Space-efficient optical computing with an integrated chip diffractive neural network. *Nat. Commun.* **13**, 1044 (2022).
61. Fu, T. *et al.* Photonic machine learning with on-chip diffractive optics. *Nat. Commun.* **14**, 70 (2023).
62. Cheng, J. *et al.* Multimodal deep learning using on-chip diffractive optics with in situ training capability. *Nat. Commun.* **15**, 6189 (2024).
63. Wu, T., Menarini, M., Gao, Z. & Feng, L. Lithography-free reconfigurable integrated photonic processor. *Nat. Photonics* **17**, 710–716 (2023).
64. Zarei, S. Programmable diffractive deep neural networks enabled by integrated rewritable metasurfaces. *Sci. Rep.* **15**, 35624 (2025).
65. Xu, Z. *et al.* Large-scale photonic chiplet taichi empowers 160-TOPS/W artificial general intelligence. *Science* **384**, 202–209 (2024).
66. Little, B. E., Chu, S. T., Haus, H. A., Foresi, J. & Laine, J.-P. Microring resonator channel dropping filters. *J. Light. Technol.* **15**, 998–1005 (1997).
67. Yariv, A. Universal relations for coupling of optical power between microresonators and dielectric waveguides. *Electron. Lett.* **36**, 321–322 (2000).
68. Bogaerts, W. *et al.* Silicon microring resonators. *Laser Photonics Rev.* **6**, 47–73 (2012).
69. Tait, A. N. *et al.* Microring weight banks. *IEEE J. Sel. Top. Quantum Electron.* **22**, 312–325 (2016).
70. Tait, A. N. *et al.* Neuromorphic photonic networks using silicon photonic weight banks. *Sci. Rep.* **7**, 7430 (2017).
71. Tait, A. N., Nahmias, M. A., Shastri, B. J. & Prucnal, P. R. Broadcast and weight: An integrated network for scalable photonic spike processing. *J. Light. Technol.* **32**, 4029–4041 (2014).
72. Ohno, S., Tang, R., Toprasertpong, K., Takagi, S. & Takenaka, M. Si microring resonator crossbar array for on-chip inference and training of the optical neural network. *ACS Photonics* **9**, 2614–2622 (2022).
73. Shao, L. *et al.* Integrated microwave acousto-optic frequency shifter on thin-film lithium niobate. *Opt. Express* **28**, 23728 (2020).
74. Tang, R. *et al.* Symmetric silicon microring resonator optical crossbar array for accelerated inference and training in deep learning. https://doi.org/10.48550/ARXIV.2401.16072 (2024) doi:10.48550/ARXIV.2401.16072.
75. Ríos, C. *et al.* In-memory computing on a photonic platform. *Sci. Adv.* **5**, eaau5759 (2019).
76. Meinders, E. R., Mijiritskii, A. V., van Pieterson, L. & Wuttig, M. Recording media. in *Optical Data Storage: Phase-Change Media and Recording* vol. 4 123–172 (Springer Netherlands, Dordrecht, 2006).
77. Rios, C., Hosseini, P., Wright, C. D., Bhaskaran, H. & Pernice, W. H. P. On-chip photonic memory elements employing phase-change materials. *Adv. Mater.* **26**, 1372–1377 (2014).
78. Wuttig, M., Bhaskaran, H. & Taubner, T. Phase-change materials for non-volatile photonic applications. *Nat. Photonics* **11**, 465–476 (2017).
79. Lee, S.-H., Jung, Y. & Agarwal, R. Highly scalable non-volatile and ultra-low-power phase-change nanowire memory. *Nat. Nanotechnol.* **2**, 626–630 (2007).
80. Dong, B. *et al.* Partial coherence enhances parallelized photonic computing. *Nature* **632**, 55–62 (2024).
81. Cheng, Z., Ríos, C., Pernice, W. H. P., Wright, C. D. & Bhaskaran, H. On-chip photonic synapse. *Sci. Adv.* **3**, e1700160 (2017).

82. Ashtiani, F., Geers, A. J. & Aflatouni, F. An on-chip photonic deep neural network for image classification. *Nature* **606**, 501–506 (2022).
83. Al-Qadasi, M. A., Chrostowski, L., Shastri, B. J. & Shekhar, S. Scaling up silicon photonic-based accelerators: Challenges and opportunities. *APL Photonics* **7**, 020902 (2022).
84. Mojaver, K. R., Safaee, S. M. R., Morrison, S. S. & Liboiron-Ladouceur, O. Recent advancements in mode division multiplexing for communication and computation in silicon photonics. *J. Light. Technol.* **42**, 7860–7870 (2024).
85. Dai, D., Wang, J. & Shi, Y. Silicon mode (de)multiplexer enabling high capacity photonic networks-on-chip with a single-wavelength-carrier light. *Opt. Lett.* **38**, 1422 (2013).
86. Luo, L.-W. *et al.* WDM-compatible mode-division multiplexing on a silicon chip. *Nat. Commun.* **5**, 3069 (2014).
87. Ling, Q. *et al.* On-chip optical matrix-vector multiplier based on mode division multiplexing. *Chip* **2**, 100061 (2023).
88. Ríos, C. *et al.* Integrated all-photonic non-volatile multi-level memory. *Nat. Photonics* **9**, 725–732 (2015).
89. Nakajima, M., Tanaka, K. & Hashimoto, T. Scalable reservoir computing on coherent linear photonic processor. *Commun. Phys.* **4**, 20 (2021).
90. Liu, A. *et al.* Wavelength division multiplexing based photonic integrated circuits on silicon-on-insulator platform. *IEEE J Sel. Top. Quantum Electron* **16**, 23–32 (2010).
91. Bai, Y. *et al.* Photonic multiplexing techniques for neuromorphic computing. *Nanophotonics* **12**, 795–817 (2023).
92. Feldmann, J., Youngblood, N., Wright, C. D., Bhaskaran, H. & Pernice, W. H. P. All-optical spiking neurosynaptic networks with self-learning capabilities. *Nature* **569**, 208–214 (2019).
93. Brückerhoff-Plückelmann, F. *et al.* Broadband photonic tensor core with integrated ultra-low crosstalk wavelength multiplexers. *Nanophotonics* **11**, 4063–4072 (2022).
94. Kitayama, K. *et al.* Novel frontier of photonics for data processing—photonic accelerator. *APL Photonics* **4**, 090901 (2019).
95. Xu, X. *et al.* 11 TOPS photonic convolutional accelerator for optical neural networks. *Nature* **589**, 44–51 (2021).
96. Sludds, A. *et al.* Delocalized photonic deep learning on the internet's edge. *Science* **378**, 270–276 (2022).
97. Wang, T. *et al.* An optical neural network using less than 1 photon per multiplication. *Nat. Commun.* **13**, 123 (2022).
98. Nahmias, M. A. *et al.* Photonic multiply-accumulate operations for neural networks. *IEEE J. Sel. Top. Quantum Electron.* **26**, 1–18 (2020).
99. Tsakyridis, A. *et al.* Photonic neural networks and optics-informed deep learning fundamentals. *APL Photonics* **9**, 011102 (2024).
100. Rahimi Kari, S., Nobile, N. A., Pantin, D., Shah, V. & Youngblood, N. Realization of an integrated coherent photonic platform for scalable matrix operations. *Optica* **11**, 542 (2024).
101. Skinner, S. R. *et al.* Reinforcement and backpropagation training for an optical neural network using self-lensing effects. *IEEE Trans. Neural Netw.* **11**, 1450–1457 (2000).
102. Xue, Z. *et al.* Fully forward mode training for optical neural networks. *Nature* **632**, 280–286 (2024).
103. Pai, S. *et al.* Experimentally realized in situ backpropagation for deep learning in photonic neural networks. *Science* **380**, 398–404 (2023).
104. Hughes, T. W., Minkov, M., Shi, Y. & Fan, S. Training of photonic neural networks through in situ backpropagation and gradient measurement. *Optica* **5**, 864 (2018).
105. Xu, T. *et al.* Control-free and efficient integrated photonic neural networks via hardware-aware training and pruning. *Optica* **11**, 1039 (2024).
106. Wang, X., Xie, P., Chen, B. & Zhang, X. Chip-based high-dimensional optical neural network. *Nano-Micro Lett.* **14**, 221 (2022).
107. Bandyopadhyay, S. *et al.* Single-chip photonic deep neural network with forward-only training. *Nat. Photonics* **18**, 1335–1343 (2024).
108. Giewont, K. *et al.* 300-mm monolithic silicon photonics foundry technology. *IEEE J. Sel. Top. Quantum Electron.* **25**, 1–11 (2019).

109. Daudlin, S. *et al.* Three-dimensional photonic integration for ultra-low-energy, high-bandwidth interchip data links. *Nat. Photonics* **19**, 502–509 (2025).
110. Atabaki, A. H. *et al.* Integrating photonics with silicon nanoelectronics for the next generation of systems on a chip. *Nature* **556**, 349–354 (2018).
111. Wan, Y. *et al.* Integrating silicon photonics with complementary metal–oxide–semiconductor technologies. *Nat. Rev. Electr. Eng.* **3**, 15–31 (2025).
112. Chao, H. *et al.* Full Poincaré polarimetry enabled through physical inference. *Optica* **9**, 1109 (2022).
113. Zhang, R. *et al.* Elliptical vectorial metrics for physically plausible polarization information analysis. *Adv. Photonics Nexus* **4**, 066015 (2025).
114. Ma, Y., Zhao, Z., Cui, J., Wang, J. & He, C. Vectorial adaptive optics for advanced imaging systems. *J. Opt.* **26**, 065402 (2024).
115. Chen, Y. *et al.* Mapping twisted light into and out of a photonic chip. *Phys. Rev. Lett.* **121**, 233602 (2018).
116. Li, M. *et al.* Printed liquid crystal optical vortex beam generators. *Adv. Opt. Mater.* **12**, 2400450 (2024).
117. Wang, A. A. *et al.* Perturbation-resilient integer arithmetic using optical skyrmions. *Nat. Photonics* **19**, 1367–1375 (2025).
118. Zhang, Y. *et al.* Skyrmions based on optical anisotropy for topological encoding. Preprint at https://doi.org/10.48550/arXiv.2508.16483 (2025).
119. He, T. *et al.* Optical skyrmions from metafibers with subwavelength features. *Nat. Commun.* **15**, 10141 (2024).
120. Xu, X. & Jin, X. Integrated photonic computing beyond the von Neumann architecture. *ACS Photonics* **10**, 1027–1036 (2023).
121. Moughames, J. *et al.* Three-dimensional waveguide interconnects for scalable integration of photonic neural networks. *Optica* **7**, 640 (2020).
122. Littlefield, A. J. *et al.* Low loss fiber-coupled volumetric interconnects fabricated via direct laser writing. *Optica* **11**, 995 (2024).
123. Cao, Z. *et al.* Programmable three-dimensional photonic neural network chip. *Nat. Commun.* https://doi.org/10.1038/s41467-026-72316-9 (2026) doi:10.1038/s41467-026-72316-9.
124. Grabulosa, A., Moughames, J., Porte, X., Kadic, M. & Brunner, D. Additive 3D photonic integration that is CMOS compatible. *Nanotechnology* **34**, 322002 (2023).
125. Skalli, A. *et al.* Photonic neuromorphic computing using vertical cavity semiconductor lasers. *Opt. Mater. Express* **12**, 2395 (2022).
126. Yeh, S. K. *et al.* Silicon photonics platform for next generation data communication technologies. in *2024 IEEE International Electron Devices Meeting (IEDM)* 1–4 (IEEE, San Francisco, CA, USA, 2024). doi:10.1109/IEDM50854.2024.10873369.
127. Xiang, C. *et al.* 3D integration enables ultralow-noise isolator-free lasers in silicon photonics. *Nature* **620**, 78–85 (2023).
128. Shekhar, S. *et al.* Roadmapping the next generation of silicon photonics. *Nat. Commun.* **15**, 751 (2024).
129. Sun, H., Qiao, Q., Guan, Q. & Zhou, G. Silicon photonic phase shifters and their applications: a review. *Micromachines* **13**, 1509 (2022).
130. Thomaschewski, M. & Bozhevolnyi, S. I. Pockels modulation in integrated nanophotonics. *Appl. Phys. Rev.* **9**, 021311 (2022).
131. Ma, L.-L. *et al.* Self-assembled liquid crystal architectures for soft matter photonics. *Light Sci. Appl.* **11**, 270 (2022).
132. Yang, Y., Forbes, A. & Cao, L. A review of liquid crystal spatial light modulators: devices and applications. *Opto-Electron. Sci.* **2**, 230026 (2023).
133. Ropač, P. *et al.* Liquid crystal 3D optical waveguides based on photoalignment. *Adv. Opt. Mater.* **13**, 2402174 (2025).
134. Seok, T. J., Quack, N., Han, S., Muller, R. S. & Wu, M. C. Large-scale broadband digital silicon photonic switches with vertical adiabatic couplers. *Optica* **3**, 64 (2016).
135. Wu, C. *et al.* Harnessing optoelectronic noises in a photonic generative network. *Sci. Adv.* **8**, eabm2956 (2022).
136. Wang, A. A. *et al.* Topological protection of optical skyrmions through complex media. *Light Sci. Appl.* **13**, 314 (2024).